\documentclass[12pt]{article}
\usepackage{epsfig}
\begin{document} 
\newcommand{\ket}[1]{|#1\rangle}
\newcommand{\bra}[1]{\langle#1|}
\newcommand{\braket}[2]{\langle#1|#2\rangle}
\newcommand{\kb}[2]{|#1\rangle\langle#2|}
\newcommand{\kbs}[3]{|#1\rangle_{#3}\phantom{i}_{#3}\langle#2|}
\newcommand{\kets}[2]{|#1\rangle_{#2}}
\newcommand{\bras}[2]{\phantom{i}_{#2}\langle#1|}
\newcommand{\af}{\alpha}
\newcommand{\bt}{\beta}
\newcommand{\gm}{\gamma}
\newcommand{\la}{\lambda}
\newcommand{\dt}{\delta}
\newcommand{\s}{\sigma}
\newcommand{\qq}{(\s_{y}\otimes\s_{y})}
\newcommand{\uu}{\rho_{12}\qq\rho_{12}^{*}\qq}
\newcommand{\tr}{\textrm{Tr}} 

\centerline{\large{Effects of cavity-field statistics on atomic entanglement
in the Jaynes-Cummings model}}

\vskip 0.2in

\centerline{Biplab Ghosh\footnote{Email: biplab@bose.res.in}, A. S. 
Majumdar\footnote{Email: archan@bose.res.in} and N. Nayak\footnote{Email: nayak@bose.res.in}}

\centerline{S. N. Bose National Centre for Basic Sciences,
Salt Lake, Kolkata 700 098, India}
\date{\today}

\vskip 0.5cm
\begin{abstract}
We study the entanglement properties of a pair of two-level 
atoms going through a cavity one after another. The initial joint state of two 
successive atoms that enter the cavity is unentangled.  Interactions mediated 
by the cavity photon field result in the final two-atom state being of a 
mixed entangled type. We consider the  field statistics
of the Fock state field, and the thermal field, respectively, 
inside the cavity.   The entanglement  of formation 
of the joint two-atom state is calculated for both these cases 
as a function of the Rabi-angle $gt$. 
We present a comparitive study of two-atom entanglement for 
low and high 
mean photon number cases corresponding to the different fields statistics.     
\end{abstract}                                                                 

\section*{I. Introduction}

Quantum entanglement has been widely observed within the framework of quantum 
optical systems such as  
cavity quantum electrodynamics. Many beautiful experiments have been 
carried out in recent years and several types of entangled states have been 
created. Practical realization of various features of quantum entanglement 
are obtained in atom-photon interactions in optical and microwave 
cavities\cite{6}. For implementation
of quantum information protocols useful in communication and 
computation\cite{5}, entanglement has to be created and preserved between
qubits that are well separated, and a recent experimental breakthrough has
been obtained by entangling two distant atomic qubits by their interaction
with the same photon\cite{8}. From the viewpoint of information
processing, quantification of entanglement is an important aspect, and
recently some studies have been 
performed to quantify the entanglement that is obtained in atom-photon 
interactions in cavities\cite{9,10,11,12}.

In the present paper we will study the dynamical generation of 
entanglement between two two-level atoms mediated by various cavity fields.
Since the atoms do not interact directly with each other, the properties
of the radiation field encountered by them bears crucially on the nature
of atomic entanglement. Our main purpose is to focus on the effect of 
different field dynamics on the magnitude of
two-atom entanglement.
The interaction between the atom and the field is governed by the
Jaynes-Cummings model\cite{13} which is experimentally realizable.
The generation of nonlocal 
correlations between the two atoms emerging from the cavity can in 
general be understood using the Horodecki theorem\cite{14}, and the joint
two-atom state is known to violate Bell-type inequalities\cite{15}.
Since the joint state of the two atoms emanating from the cavity is not a pure 
state, we quantify the entanglement using the well-known measure appropriate
for mixed states, i.e., the entanglement of formation\cite{16}.
We investigate how the statistics of different types of radiation fields
influence the quantitative dynamics of atomic entanglement.

The structure of the paper is as follows. In Section II we review briefly the 
interaction 
between two-level atom and single mode radiation field inside a cavity 
described by the Jaynes-Cummings model. 
In Section III we show how the entanglement between two spatially separated 
atoms is generated. We observe robust atom-atom entanglement mediated by the
Fock state field, and the thermal field, respectively. We demonstrate how
the various field statistics are reflected in two-atom entanglement as
a function of the average photon number of the cavity fields and the
Rabi angle. Several
distinctive characteristics of the entanglement generated by the different
fields through the Jaynes-Cummings interaction are discussed in comparison 
with some earlier results obtained for the Tavis-Cummings interaction\cite{24}.
A common feature that is observed is 
that for the cavity low photon number case, the entanglement between the 
two atoms decreases with increasing average photon number of the field. A 
summary of our results and some concluding remarks are presented in Section IV.

\section*{II. Entanglement mediated by the Jaynes-Cummings interaction}

The Jaynes-Cummings (JC) model is one of the most studied models in quantum 
optics. 
Our aim is to study the entanglement between atoms mediated by the optical
field, where the light-atom interaction is governed by the JC model.
The JC model conists of a two-level 
atom coupled to a single-mode radiation field inside a cavity. 
A two level atom is formally analogous to a spin-1/2 system. Let us denote 
the upper level of the atom as $\ket{e}$ and the lower level as $\ket{g}$ 
and the spin (atomic) raising and lowering operators can be defined 
as $\s^+=\kb{e}{g}$ and $\s^-=\kb{g}{e}$, respectively,
with the commutation relation 
\begin{eqnarray}
[\s^{+},\s^{-}]=\kb{e}{e}-\kb{g}{g}=\s_{z}.
\label{1}
\end{eqnarray}

A quantum mechanical field can be represented as (for the present purpose, we
consider a single mode field)
\begin{eqnarray} 
E(t)=\frac{1}{2}[a e^{-i\omega t} + a^\dagger e^{i\omega t}]
\label{2}
\end{eqnarray} 
Here $a$ and $a^\dagger$ are annihilation and creation 
operators, respectively. The graininess of the radiation field is represented 
by the photon number state $\ket{n}$, $n=0, 1, 2, ..... $, such that 
$a\ket{n}=\sqrt{n}\ket{n-1}$ and $a^\dagger\ket{n}=\sqrt{n+1}\ket{n+1}$.
It is an eigenstate of the number operator $\hat{n}=a^\dagger a$ 
given by $\hat{n}\ket{n}=n\ket{n}$.
The field in Eq.(2) can be represented by a quantum mechanical state vector 
$\ket{\psi}$ which is a linear superposition of of the number states 
$\ket{n}$, i.e., $\ket{\psi}=\sum_{n=0}^\infty c_n\ket{n}$.
where $c_n$ is, in general, complex and gives the probabilty of the field 
having
$n$ photons by the relation $P_n=\braket{n}{\psi}\braket{\psi}{n}=|c_n|^2$.
The field obeys quantum statistics and its average photon number is given 
by 
\begin{eqnarray}
<n>=\sum_{n=0}^\infty nP_n
\label{6}
\end{eqnarray}
with the intensity of the field $I\propto<n>$. The statistics brings in a 
quantum mechanical noise which is represented by the variance 
\begin{eqnarray}
V=\frac{<n^2>-<n>}{<n^2>}.
\label{7}
\end{eqnarray}
$V=1$ is for coherent state field and $V<1$ signifies a non-classical state. 
The parameters $<n>$ and $V$ give a fair description of the quantum 
mechanical nature of the radiation field. 

The interaction picture Hamiltonian of the joint atom-field system can be 
written in the rotating 
wave approximation \cite{19} as, 
\begin{eqnarray}
H_{I}=g(\s^+ a+\s^-a^\dagger).
\label{10}
\end{eqnarray}
where $a^\dagger$ and  $a$ are usual creation and destruction operators of the
radiation field.
Here we have considered the quality factor of the cavity $Q=\infty$
since the cavity-QED related experiments are carried out with cavities 
with very high $Q$ \cite{6}. We shall consider the cavity field to 
be in a Fock and a 
thermal state, respectively.

With the passage of the two atoms, one after the other, the joint state
of both the atoms and the field at some instance $t$ may be denoted
by $|\Psi(t)>_{a-a-f}$. 
The corresponding atom-atom-field pure density state is
$\rho(t)=\kb{\Psi(t)}{\Psi(t)}$.
In order to quantify the entanglement between the two atoms, the field
variables have to be traced out.
The reduced mixed density state of two atoms after taking trace over the 
field is
\begin{eqnarray} 
\rho(t)={\textrm Tr}_{\mathrm{field}}(|\Psi(t)>_{a-a-f.a-a-f}<\Psi(t)|)
\label{12}
\end{eqnarray}
Entanglement within pure states of bipartite system can be measured 
by the Von Neumann entropy of the reduced density matrices. For mixed states
the entanglement of formation\cite{16} is a widely used measure
for computing atomic entagelement in quantum optical 
systems\cite{8,9,10,11,12}.
The entanglement of formation for a bipartite density operator 
$\rho$ is given by 
\begin{eqnarray}
E_{F}(\rho)=h\left(\frac{1+\sqrt{1-C^{2}(\rho)}}{2}\right),
\label{13}
\end{eqnarray}
where $C$ is called the concurrence defined as
\begin{eqnarray}
C(\rho)=\max(0, \sqrt\la_1-\sqrt\la_2-\sqrt\la_3-\sqrt\la_4),
\label{14}
\end{eqnarray}
where the 
$\la_{i}$ are the  eigenvalues of $\uu$ in descending order,
is the binary entropy function.
The entanglement of formation is a monotone of the
concurrence. 

\section*{III. Entanglement features of cavity fields}

\begin{figure}[h!]
\begin{center}
\includegraphics[width=6cm]{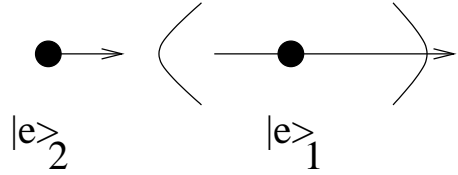}
\caption{Two atoms prepared in excited states are pass through a single 
mode cavity one after the other.}
\end{center}
\end{figure}

We consider a micromaser  system in which atoms are sent 
into the cavity at such a rate that the probability of two atoms being 
present there is negligibly small.  Our purpose here is to 
show the influence 
of the photon statistics of the driving fields (radiation field with which the 
atoms interact) on atomic entanglement. 
For this sake, we consider the cavity to be of a non-leaky type, that is, 
$Q=\infty$. In fact, the cavity-QED experiments are very close to such  
situations\cite{6}.
In the following, we consider two different kinds of radiation fields,
i.e., the Fock state field, and the thermal field, respectively. 

\subsection*{A. FOCK STATE FIELD}

A Fock state is written as $\ket{n}$ with $n$ an integer value, signifying
that there are $n$ quanta of excitation in the mode. 
$\ket{0}$ corresponds to the ground state (no excitation).
Fock states form the most convenient basis of the Fock space. 
The amplitudes $c_n$  obey the delta function relation
$c_n=\delta_{m,n}$
where $m$ is the photon number of the Fock state. The variance is given by
$V=1-\frac{1}{m}$.
So, for small values of $m$, $V<1$ and field has non-classical character. 
For large $m$, $V$ tends towards the classical limit. This feature is 
reflected in the entanglement generated between the two atoms, as we shall
see later.

Let us first consider the passage of the first atom, initially in the excited
state $|e>$,  through the cavity. The joint atom-field state is given by
\begin{eqnarray}
\ket{\Psi(t=0)}_{a-f}=\ket{e}\otimes\ket{n}.
\label{18}
\end{eqnarray}
The atom-field wave function evolves with the interaction given by Eq.(5) 
to
\begin{eqnarray}
\ket{\Psi(t)_{a-f}}=\cos{(\sqrt{n+1}gt)}\ket{e,n}
+\sin{(\sqrt{n+1}gt)}\ket{g,n+1}
\label{19}
\end{eqnarray}
The next atom which enters the cavity 
interacts with this ``changed" field and thus a correlation develops 
between the two atoms via the cavity field.
The joint state of the two atoms and the field is given by 
\begin{eqnarray}
\ket{\Psi(t)}_{a-a-f}=
\alpha_1\ket{e_1,e_2,n}+\alpha_2\ket{e_1,g_2,n+1}
+\alpha_3\ket{g_1,e_2,n+1}+\alpha_4\ket{g_1,g_2,n+2}
\label{20}
\end{eqnarray} 
where
\begin{eqnarray}
\alpha_1=\cos^2{(\sqrt{n+1}gt)}, \>\>
\alpha_2=\cos{(\sqrt{n+1}gt)}\sin{(\sqrt{n+1}gt)}, \nonumber\\
\alpha_3=\cos{(\sqrt{n+2}gt)}\sin{(\sqrt{n+1}gt)},\>\>
\alpha_4=\sin{(\sqrt{n+1}gt)}\sin{(\sqrt{n+2}gt)}.
\label{21}
\end{eqnarray} 
The reduced mixed density state of two atoms after tracing over the field is
given by (we display the non-vanishing terms only)
\begin{eqnarray} 
\rho(t)_{a-a}=\textrm{tr}_{f}
(\rho(t)_{a-a-f})
=\alpha_1^2\kb{e_1e_2}{e_1e_2}+\alpha_2^2\kb{e_1g_2}{e_1g_2}\nonumber\\
+\alpha_3^2\kb{g_1e_2}{g_1e_2}+\alpha_2\alpha_3\kb{e_1g_2}{g_1e_2}
+\alpha_2\alpha_3\kb{g_1e_2}{e_1g_2}+\alpha_4^2\kb{g_1g_2}{g_1g_2}.
\label{23}
\end{eqnarray}
We compute the entanglement of formation $E_f$  for this
bipartite two-atom state. In Figure~2 
$E_F$ is plotted versus the Rabi angle $gt$ for
different values of $n$. 
\vskip 1cm

\begin{figure}[h!]
\begin{center}
\includegraphics[width=8cm]{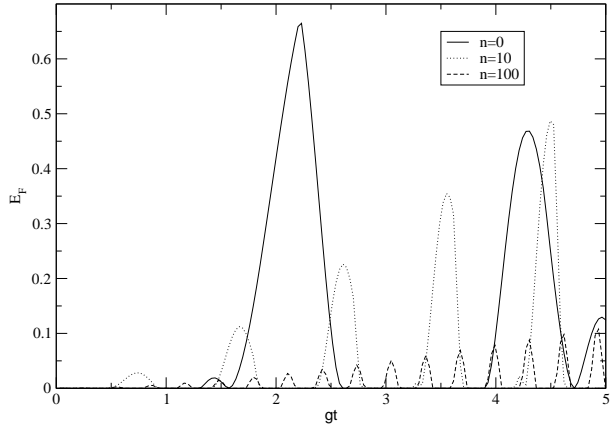}
\caption{Atom-atom entanglement versus $gt$. Solid line, dotted line, 
and dashed line
 indicate $E_F$ between two atoms when the cavity fock states are
$n=0$, $n=10$, and
$n=100$ respectively.}
\end{center}
\end{figure}

The peaks of the entanglement of formation are reflective of the photon
statistics that are typical in micromaser dynamics\cite{23}.
We see that $E_F$ falls off sharply as $n$ increases. The 
non-classical character of the field for small values of the average 
photon number
$n$, is reflected in larger entanglement
between the two atoms. An interesting comparison can be made with the case
of the Tavis-Cummings model\cite{24} which is employed when two atoms are
present simultaneously inside the cavity.
Although simultaneous interaction of two excited atoms with 
Fock state field never results in two-atom entanglement as was shown by 
Tessier et al.\cite{12}, the notable difference here is that in the JC
dynamics modelling the micromaser
one always gets two-atom 
entanglement mediated by the Fock state cavity field, as we see in Figure~2.

\subsection*{B. THERMAL FIELD}

The thermal field is the most easily available radiation field, and so, 
its influence on the 
entanglement of spins is of much interest. The field at thermal 
equilibrium obeying Bose-Einstein statistics has an average photon number 
at temperature $T^0 K$, given by 
\begin{eqnarray}
<n>=\frac{1}{e^{\hbar \omega/kT}-1}.
\label{24}
\end{eqnarray}
The photon statistics is governed by the distribution $P_n$ given by
\begin{eqnarray}
P_n=\frac{<n>^n}{(1+<n>)^{n+1}}.
\label{25}
\end{eqnarray}
This distribution function always peaks at zero, i.e., $n_{peak}=0$.
For a thermal field distribution function for the cavity field, the joint  
two-atom-cavity state is obtained by summing over all $n$, and is given by 
\begin{eqnarray}
\ket{\Psi(t)}_{a-a-f}=\sum_nA_n[\cos^2{(\sqrt{n+1}gt)}
\ket{e_1,e_2,n}
+\cos{(\sqrt{n+1}gt)}\sin{(\sqrt{n+1}gt)}\ket{e_1,g_2,n+1}\nonumber\\
+\cos{(\sqrt{n+2}gt)}\sin{(\sqrt{n+1}gt)}\ket{g_1,e_2,n+1}
+\sin{(\sqrt{n+1}gt)}\sin{(\sqrt{n+2}gt)}\ket{g_1,g_2,n+2}]
\label{26}
\end{eqnarray} 
where $P_n=|A_n|^2$ is the photon distribution function of the thermal field.

\vskip 1cm

\begin{figure}[h!]
\begin{center}
\includegraphics[width=8cm]{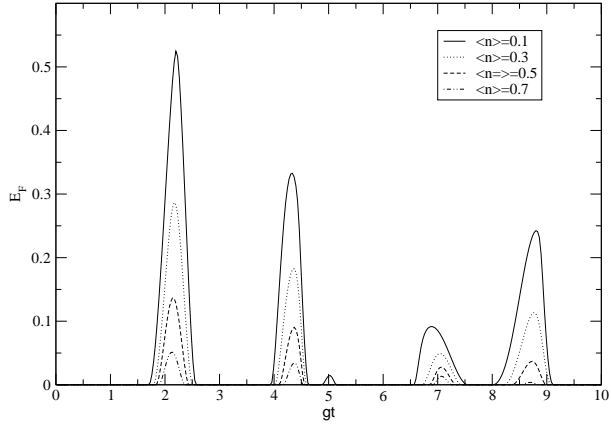}
\caption{Atom-atom entanglement of formation mediated by the thermal 
cavity field
is plotted versus $gt$.}
\end{center}
\end{figure}

The reduced mixed density state of two atoms after passing through the 
the thermal cavity field can be written as
\begin{eqnarray} 
\rho(t)_{a-a}=\textrm{tr}_{f}
(\rho(t)_{a-a-f})=
\beta_1\kb{e_1e_2}{e_1e_2}+\beta_2\kb{e_1g_2}{e_1g_2}\nonumber\\
+\beta_3\kb{g_1e_2}{g_1e_2}+\beta_4\kb{e_1g_2}{g_1e_2}
+\beta_4\kb{g_1e_2}{e_1g_2}+\beta_5\kb{g_1g_2}{g_1g_2},
\label{27}
\end{eqnarray}
where
\begin{eqnarray}
\beta_1=\sum_nP_n\cos^4{(\sqrt{n+1}gt)}, \>
\beta_2= \sum_nP_n\cos^2{(\sqrt{n+1}gt)}\times
\sin^2{(\sqrt{n+1}gt)},\nonumber
\end{eqnarray}
\begin{eqnarray}
\beta_3=\sum_nP_n\cos^2{(\sqrt{n+2}gt)}\times
\sin^2{(\sqrt{n+1}gt)},\>
  \beta_5=\sum_nP_n\sin^2{(\sqrt{n+1}gt)}\times
\sin^2{(\sqrt{n+2}gt)},\nonumber
\end{eqnarray}
\begin{eqnarray}
\beta_4=\sum_nP_n\sin^2{(\sqrt{n+1}gt)}\times
\cos{(\sqrt{n+1}gt)}\cos{(\sqrt{n+2}gt)}.
\label{28}
\end{eqnarray}

We compute the entanglement of formation $E_F$ for the above two-atom 
state and plot it versus the Rabi angle $gt$ for 
different values of average photon number $<n>$ in Figure~3.
It is interesting to note that the
thermal field which has miminimal information can nevertheless 
entangle qubits that are prepared initially in a separable state.
In the context of the Tavis-Cummings framework when both the atoms interact
simultaneously with the radiation field, Kim et al. \cite{11} 
have noticed similar trends in the entanglement mediated by the thermal
field. Thus both Jaynes-Cummings and the Tavis-Cummings models of
atom-photon interaction
generate similar entanglement when the radiation field is thermal, whereas
for the Fock state field case the situation is contrasting as observed 
earlier.

\section*{IV. Conclusions}

To summarize, in this paper we have presented a 
realistic micromaser-type model where two 
spatially separated atoms are entangled via a cavity field. 
The entanglement between the two separate atoms builds up via atom-photon 
interactions inside the cavity, even though no single atom interacts 
directly with another. We have computed the two-atom entanglement as
measured by the entanglement of formation $E_F$, for the case of two
different types of radiation fields, i.e., the Fock state field and the thermal
field, respectively. Our purpose has
been to study the effects of the statistics of the bosonic radiation field
on the dynamics of the entanglement of two atomic qubits, i.e., two 
fermionic systems. Several interesting features of atomic entanglement 
are observed.

We first show that for the Fock state cavity field, entanglement between
two successively passing atoms can be generated as a consequence of
Jaynes-Cummings (JC) dynamics. This is in contrast to the case when both the
atoms reside together inside the cavity when Tavis-Cummings (TC) dynamics
for atom-photon interactions is unable to generate atomic 
entanglement\cite{12}. We then study the entanglement mediated by the thermal 
radiation field. It is interesting to note that the thermal field
which carries minimum information is still able to produce atomic entanglement
through both the JC interaction as seen here, and also through the
TC interaction as was observed earlier\cite{11}. However, the thermal field
having a high value of the average photon number loses its ability to entangle
atomic qubits passing through it. 

Finally, we would like to reemphasize that the quantitative study
of entanglement produced in various types of atom-photon interactions 
is a relevant
arena for investigations. Atom-photon interactions and the generation
of entanglement mediated through them are expected to play an important
role in possible future practical realizatons in the field of quantum
communications\cite{8,26}.  
Recently,  the possibility of entanglement of a thermal radiation 
field with high temperature phonons associated with moving mirrors of a 
cavity has been shown\cite{28}, brightening the prospects for creating
macroscopic entanglement. Even from a purely pedagogical perspective, 
investigations of quantitative entanglement in atom-photon interactions could
lead to interesting insights on the curious properties of
entanglement such as its `monogamous' nature\cite{29}.

\end{document}